# A piezo-metastructure with bistable circuit shunts for adaptive nonreciprocal wave transmission

Yisheng Zheng*[1,2], Zhen Wu[2], Xinong Zhang[1], K.W. Wang[2]
[1]State Key Laboratory for Strength and Vibration of Mechanical Structures, Xi'an Jiaotong University, Xi'an 710049, China
[2]Department of Mechanical Engineering, University of Michigan, Ann Arbor, MI 48109, USA
*Corresponding author: zhysh.966@stu.xjtu.edu.cn

**Abstract**

In this paper, we present a piezo-metastructure shunted with bistable circuits to achieve adaptive nonreciprocal elastic wave transmission. Static properties of the bistable circuit are first investigated, followed by numerical investigation of wave transmission characteristics on the nonlinear piezo-metastructure. Both the local-resonance bandgap and the bandgap transmission phenomenon, also known as *supratransmission*, are explored and investigated. By introducing circuit asymmetry, the supratransmission thresholds, critical excitation amplitudes to enable bandgap transmission, are found at different levels for the different wave transmission directions and hence creating an excitation amplitude range within which nonreciprocal wave transmission can be facilitated. Effect of the asymmetry factor on nonreciprocity properties is subsequently analyzed and the trade-offs between the forward transmission amplitude and the range of excitation with nonreciprocity are identified. Additionally, it is demonstrated that wave transmission characteristics of the proposed nonlinear piezo-metastructure can be adaptively tuned by conveniently adjusting stable equilibria of the bistable circuit. Lastly, the observed wave transmission properties are further corroborated by experimental investigations. Overall, the results illustrate a novel means to manipulate unidirectional elastic wave transmission using a nonlinear piezo-metastructure.

## 1. Introduction

For conventional linear time-invariant systems, wave transmission has to obey the reciprocity theorem [1] and hence if wave can propagate in one direction, it can also propagate in the reverse direction. The electric diode however is a nonreciprocal system that allows current to transmit in only one direction. Because of this special phenomenon, electric diodes have been widely used in circuits and greatly advanced the electronic technology. Inspired by electric diodes, the nonreciprocal wave transmission phenomenon has attracted much attention in the area of electromagnetic wave [2], thermal wave [3] and acoustic/elastic wave [4]. For acoustic/elastic wave, which is the focus of this paper, some relevant investigations are reviewed in the following. Fleury et al. [5] proposed an acoustic circulator using circulating fluid to bias the resonant ring cavity and imitating the Zeeman Effect of nonreciprocal electromagnetic systems. Nassar et al. [6] and Trainiti et al. [7] explored time-variant systems with modulated density and modulus in space and time domain to yield nonreciprocal wave transmission. These abovementioned methods are based on linear systems with active modulation. Another promising approach to achieve nonreciprocity is to utilize nonlinear metamaterials, and many nonlinear nonreciprocal metamaterials have been proposed and studied. For instance, Liang et al. [4,8] proposed a nonlinear acoustic metamaterial composed of a superlattice and a strongly nonlinear medium; Boechler et al. [9] investigated a bifurcation-based acoustic rectifier with rectification ratio greater than $10^4$; Liu et al. [10] proposed using an amplitude adjustment element on one end of a weakly nonlinear chain to shift its bandgap in one direction and hence created a nonreciprocal frequency range, which was further experimentally verified by Cui et al. [11]; Lepri and Casati [12] considered a layered nonlinear, nonmirror-symmetric model with spatially varying coefficients embedded in an linear lattice and found that nonreciprocity is associated with the shift of nonlinear resonances; Wu et al. [13,14] presented a metastructure composed of metastable elements and exploited using supratransmission to realize nonreciprocal wave transmission. However, to the best of the authors' knowledge, previous investigations are all relying on mechanical metamaterials, which are cumbersome to fabricate and integrate physically. Also for most of the approaches, the nonlinear characteristics of the constituents cannot be easily adjusted once they are fabricated and as a result the wave transmission properties of the system cannot be conveniently adjusted.

Piezoelectric material-systems, which have the benefit of light add-on mass and easy implementation, have been extensively studied for various applications [15]. More specifically, linear piezo-metamaterials have recently been investigated and exploited for manipulating wave propagation [16–20]. However, little attention has been paid to piezo-metamaterials that are shunted with nonlinear circuits [21], which can be potentially designed to realize functionalities that have been achieved using nonlinear mechanical metamaterials, such as acoustic rectifier [8], logic element [22], phononic metasurface [23] and phononic transistor [24]. Since nonlinearity is induced by circuits, it is expected that the nonlinear piezo-metamaterial can be easily realized using the circuit integrating and fabrication technology [25]. Compared with pure mechanical systems, the nonlinear properties of circuits can be much more conveniently adjusted. Therefore, we seek to explore a piezo-metastructure shunted with nonlinear circuits, bistable circuits [26,27] specifically in this paper, to realize adaptive nonreciprocal wave transmission. The proposed bistable circuit owns the characteristic that its stable equilibria can be conveniently adjusted, which can be exploited to realize programmable wave transmission properties.

The paper is organized as follows. In Section 2, the model of the nonlinear piezo-metastructure is presented and its dynamic equation is derived. Numerical investigations are performed to elucidate the bandgap, supratransmission and nonreciprocity of the proposed system, and discussed in Section 3. In Section 4, experiments conducted to verify the observed wave transmission properties are presented. Lastly, some concluding remarks are summarized in Section 5.

## 2. The piezo-metastructure with bistable circuit shunts

The schematic of the proposed system is shown in Fig. 1(a), which is a clamped-clamped piezoelectric beam composed of a substrate layer and two piezoelectric layers. Fig. 1(b) is the zoom in view of a beam segment illustrating geometric parameters. The electrodes are placed periodically on both sides of the beam and the thicknesses of which are neglected. In part II of the system, as denoted in Fig. 1(a), the electrodes on both sides of each unit are shunted with bistable circuits. To intentionally induce asymmetry, the electrodes in parts I and III are shunted with $L$-$R$ circuits and short circuits, respectively. With the existence of the piezoelectric transducer capacitance, the equivalent $L$-$C$-$R$ circuits are utilized for analysis (shown to be the linear circuits in part I of Fig. 1(a)). The force excitation $F$ is imposed on point $A_1$ (or $A_2$) of the beam and the velocity response of points $S_1$ and $S_2$ are evaluated to determine the wave transmission properties. The wave transmission direction is defined as *backward* if $F$ is imposed on point $A_1$ and *forward* if $F$ is imposed on point $A_2$.

The bistable circuit used in part II is shown in Fig. 1(c). The nonlinearity of the circuit is due to the existence of diodes. The voltage source $V_a$ applied over the diodes can be used to adjust stable equilibria of the bistable circuit. Two op-amps with negative feedback are employed in the bistable circuit. $C_p$ is the capacitance of the piezoelectric transducer. The linear circuit in part I is a series of a resistor $R_a$ and an inductor $L_a$. The inductance $L_a$ is chosen to be such that the resonance frequency of the linear circuit equals to that of the linearized bistable circuit operating at one of its stable equilibria. Since electrodes in part III are shorted, they have no effect on wave transmission. When all bistable circuits operate near the stable equilibria, the proposed system is equivalent to a linear piezo-metamaterial shunted with $L$-$R$ circuits, and it exhibits a local-resonance bandgap [16,28]. However, when large excitations are applied to the system, response will be primarily dominated by the nonlinear effect of the bistable circuits.

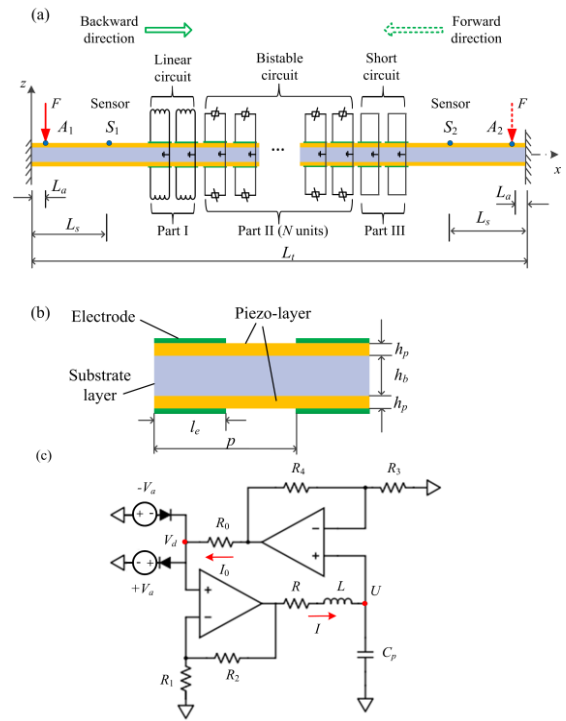

**Fig. 1** (a) Schematic of the nonlinear piezo-metastructure; (b) zoom in view of a beam segment; (c) the bistable circuit diagram

## 2.1 The bistable circuit

Before analyzing the response of the overall system, we first study properties of the bistable circuit. The dynamic equation of the bistable circuit without excitations can be derived as

$$\begin{cases} C_p \dfrac{dU}{dt} - I = 0 \\ L \dfrac{dI}{dt} + RI + P(U) = 0 \end{cases} \quad (1)$$

where the voltage response $U$, indicated in Fig. 1(c), describes the voltage on the piezoelectric transducer and could affect the dynamic response of the mechanical structure. Eq.(1) can be further transformed to

$$LC_p \ddot{U} + RC_p \dot{U} + P(U) = 0 \quad (2)$$

where the voltage function $P(U)$ is

$$P(U) = U - gV_d \quad (3)$$

In Eq. (3), $g$ is the amplification factor of one op-amp with negative feedback, given as $g = 1 + R_2/R_1$.

It can be seen from Eq. (2) that the dynamic equation of the bistable circuit is equivalent to that of a mechanical oscillator: $U$ is equivalent to the displacement; the voltage function $P(U)$ is equivalent to the restoring force; hence the equilibrium voltage $U_e$ of the bistable circuit can be obtained by solving $P(U)=0$.

Using the ideal diode model [27,29], the voltage $V_d$ on the diode node can be determined as

$$V_d = \begin{cases} V_a + \text{VD}, & g_1 U > V_a + \text{VD} \\ g_1 U, & -(V_a + \text{VD}) \le g_1 U \le V_a + \text{VD} \\ -(V_a + \text{VD}), & g_1 U < -(V_a + \text{VD}) \end{cases} \quad (4)$$

where the op-amp amplification factor $g_1$ is defined as $g_1 = 1 + R_4/R_3$ and VD is the constant diode forward voltage beyond which the current can transmit through the diode. Substituting Eq. (4) into Eq. (3), it is shown that the voltage function $P(U)$ is a piecewise linear function and the bistable circuit working near the stable equilibria is equivalent to a linear $L$-$C$-$R$ circuit. This simplified model can serve as a convenient tool to qualitatively estimate behaviours of the bistable circuit. However, for a practical diode, the diode forward voltage VD varies with the static current flowing through it. As a result, the bistable circuit operating around the stable equilibria is not linear and its linearized resonance frequency at the stable equilibria will deviate from that of the linear circuit, i.e. $\sqrt{1/LC_p}/(2\pi)$. This kind of characteristic cannot be reflected using the ideal diode model. Hence, in this paper, we will use the Shockley diode model for more accurate characterization [30], based on which, the current $I_0$ shown in Fig. 1(c) can be derived as

$$I_0 = I_S[e^{(V_d - V_a)/\kappa/V_T} - 1] - I_S[e^{(-V_d - V_a)/\kappa/V_T} - 1] \quad (5)$$

where $I_S$, $V_T$ and $\kappa$ are diode model constants. On the other hand, $I_0$ can be expressed as

$$I_0 = (g_1 U - V_d)/R_0 \quad (6)$$

Hence, the relation between $U$ and $V_d$ can be derived by combining Eqs.(5) and (6) as

$$g_1 U = V_d + 2R_0 I_S e^{(-V_a/\kappa/V_T)} \sinh\left(\dfrac{V_d}{\kappa V_T}\right) \quad (7)$$

From Eqs.(3) and (7), the voltage function $P(U)$ can be obtained, which can be used to characterize static properties of the bistable circuit. It can also be derived from Eqs. (3) and (7) that

$$g_1 P(U) = [(V_d + 2R_0 I_S e^{(-V_a/\kappa/V_T)} \sinh\left(\dfrac{V_d}{\kappa V_T}\right)] - gg_1 V_d) \quad (8)$$

When $P(U)=0$, which means that the bistable circuit is at its static equilibrium, $V_d$ can be obtained from Eq. (8) denoted as $V_{de}$. Eq. (8) implies that $V_{de}$ can be affected by two parameters: $V_a$ and $gg_1$. Since the equilibrium of $U$ can be obtained from Eq. (3) as

$$U_e = g \cdot V_{de} \quad (9)$$

we can induce that $U_e$ also depends on the voltage source applied on diode $V_a$ and the op-amp amplification factors $g$ and $g_1$. Linearizing the nonlinear relation between $U$ and $V_d$ in Eq. (7) around $V_{de}$, it can be expressed as

$$g_1 U = f(V_a, V_{de}) \cdot V_d + h(V_a, V_{de}) \quad (10)$$

where $f(V_a, V_{de})$ and $h(V_a, V_{de})$ are two functions of variables $V_a$ and $V_{de}$. $f(V_a, V_{de})$ can be derived from Eq.(7) as

$$f(V_a, V_{de}) = 1 + \dfrac{2R_0 I_S}{\kappa V_T} e^{(-V_a/\kappa/V_T)} \cosh\left(\dfrac{V_{de}}{\kappa V_T}\right) \quad (11)$$

From Eqs.(2)(3)(10), the linearized resonance frequency of the bistable circuit around its stable equilibrium is

$$f_n = \dfrac{1}{2\pi}\sqrt{[1 - \dfrac{gg_1}{f(V_a, V_{de})}]\dfrac{1}{LC_p}} \quad (12)$$

which shows that $f_n$ depends on the controlled DC voltage source $V_a$ and the product of op-amp amplification factors $gg_1$. Eqs. (11) and (12) also indicate that $f_n$ is always smaller than that of the corresponding linear *L-C-R* circuit, which is $\sqrt{1/LC_p}/(2\pi)$.

The parameters of the bistable circuit are shown in Table 1. The profiles of voltage function $P(U)$ for different $V_a$ are depicted in Fig. 2. It can be seen that for cases $V_a$=-0.2 V, 0 V, and 0.2 V, there exists two stable equilibria and one unstable equilibrium as well as the negative-stiffness range, where the slope of the voltage function curve is negative. These are the static properties of bistable oscillators [31] and hence it verifies that the proposed circuit has the bistability property. Fig. 2 also shows that, by decreasing $V_a$ from 0.2 V to -0.2 V, the distance between the two stable equilibria as well as the negative-stiffness range are reduced. Furthermore, if $V_a$ is decreased to $V_a$=-0.4 V, the negative-stiffness range disappears and the circuit degenerates into a monostable system. Using Eqs. (8) and (9), the stable equilibrium $U_e$ of the bistable circuit can further be obtained analytically and its relation with $V_a$ in the range -0.3 V ~ 0.6 V is presented in Fig. 3, which shows that the relation between $U_e$ and $V_a$ is quasi-linear, which can be approximated as $U_e = g \cdot (V_a + \text{VD})$ using the ideal diode model. Overall, the results shown in Fig. 2 and 3 demonstrate the ability of utilizing DC voltage sources $V_a$ to adjust stable equilibria of the bistable circuit.

Other than the DC voltage source $V_a$, the op-amps amplification factors $g$ and $g_1$, which depend on $R_1$, $R_2$, $R_3$ and $R_4$, can also affect the profile of voltage function $P(U)$. Especially, varying $g$ and $g_1$ would affect the ratio $r_n$ of negative stiffness range to the distance between the two stable equilibria, which can be derived as $r_n = 1/(gg_1)$ using the ideal diode model Eq. (3) and (4). Hence, the ratio $r_n$ would keep unchanged if $gg_1$ is constant. If use the Shockley diode model, the value of $r_n$ is slightly different but would not change the qualitative conclusion. As shown in Fig. 4, we set $V_a$=0 and keep $gg_1$ unchanged, the voltage functions for cases with different $g$ and $g_1$ are obtained. When $g$ is increased (accordingly, $g_1$ is decreased), the negative stiffness range and the distance between the two stable equilibria are increased simultaneously.

Therefore, the nonlinear properties of the proposed bistable circuit, which is governed by the voltage function $P(U)$ and characterized by the stable equilibria $U_e$, can be adaptive by adjusting the DC voltages source $V_a$ and the op-amps amplification factors ($g$ and $g_1$). This provides us the opportunities to adjust the wave transmission properties of the proposed nonlinear piezo-metastructure.

**Table 1** Parameters of the bistable circuit in simulation

| Parameters | Value |
|---|---|
| $L$ | 0.1 [H] |
| $C_p$ | 10.7 [nF] |
| $R$ | 87 [Ω] |
| $R_0$ | 15 [KΩ] |
| $I_S$ | $2.52 \times 10^{-9}$ [A] |
| $V_T$ | $2.6 \times 10^{-2}$ [V] |
| $\kappa$ | 1.752 |

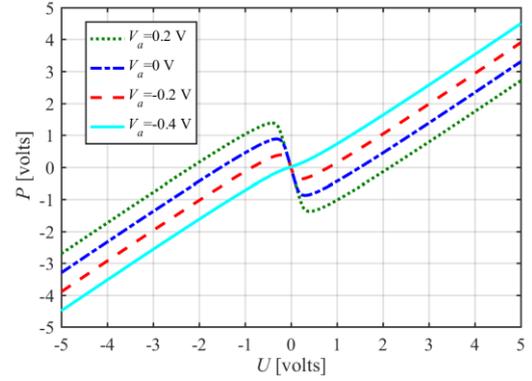

**Fig. 2** Effect of $V_a$ on the voltage function $P(U)$ when $g$=3, $g_1$=2.

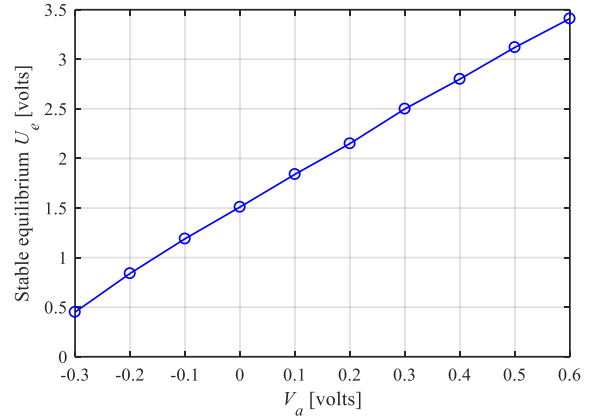

**Fig. 3** Stable equilibria of the bistable circuit with different $V_a$ controlled by DC voltage sources when $g$=3, $g_1$=2.

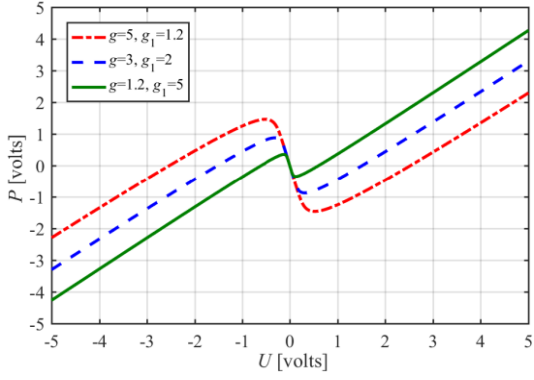

**Fig. 4** Effect of $g$ and $g_1$ on the voltage function $P(U)$ when $V_a=0$ V.

*2.2 Dynamic model of the piezo-metastructure*

With the derived dynamic equation of the bistable circuit, in this section, we seek to determine the equations of motion of the overall system. Following the method used for a linear piezo-metastructure [19], we derive the discretized dynamic equations with the proposed nonlinear piezoelectric system.

To investigate the bandgap and supratransmission properties, the linear circuits of part I, as shown in Fig. 1(a), are shorted and the system becomes symmetric. In this case, the dynamic equation of the clamped-clamped beam is expressed as

$$\begin{cases} EI\dfrac{\partial^4 w}{\partial x^4}+m\dfrac{\partial^2 w}{\partial t^2}-2\lambda\sum_{i=1}^{N}U_i\dfrac{d^2}{dx^2}[H(x-x_{Li})-H(x-x_{Ri})] \\ = F\cdot\delta(x-x_F) \\ w|_{x=0}=0,\ \dfrac{\partial w}{\partial x}|_{x=0}=0,\ w|_{x=L_t}=0,\ \dfrac{\partial w}{\partial x}|_{x=L_t}=0 \\ i=1,2...N \end{cases} \quad (13)$$

where

$$\lambda = \dfrac{d_{31}w_p}{2s_{11}^E h_p}[(h_p+\dfrac{h_b}{2})^2-\dfrac{h_b^2}{4}] \quad (14)$$

and it represents the mechanical-electric coupling factor. $EI$ is the bending stiffness of the piezo-beam, given as

$$EI = \dfrac{E_b w_b h_b^3}{12}+\dfrac{E_p w_p[(h_b+2h_p)^3-h_b^3]}{12} \quad (15)$$

$m$ is the mass per unit length:

$$m = \rho_b w_b h_b + 2\rho_p w_p h_p \quad (16)$$

$w$ is the transverse displacement; $x_{Li}$ and $x_{Ri}$ are end positions of eletrodes in the $i^{th}$ unit on the beam, and the electrode length is $x_{Ri}-x_{Li}=l_e$; $x_F$ is the position of the force excitation; $w_b$, $h_b$, $E_b$ and $\rho_b$ are the width, height, modulus and density of the substrate layer, respectively; $w_p$, $h_p$, $E_p$ and $\rho_p$ are the width, height, modulus and density of the piezoelectric layer, respectively; $H(x)$ is the Heaviside function; $d_{31}$ is the piezoelectric strain constant; $s_{11}^E$ is the elastic compliance of piezoelectric layers at constant electric field $s_{11}^E = 1/E_p$; $N$ is the unit number in part II of the system.

Since the response of the two bistable circuits within the same unit is identical, we only need to establish the dynamic equation of one circuit for each unit, which can be given as

$$\begin{cases} C_{pi}\dfrac{dU_i}{dt}-I_i+\lambda\int_{x_{Li}}^{x_{Ri}}\dfrac{\partial^3 w}{\partial x^2 \partial t}dx=0 \\ L\dfrac{dI_i}{dt}+RI_i+P(U_i)=0 \end{cases},\quad i=1,2...N \quad (17)$$

where

$$C_{pi}=(\varepsilon_{33}^T-\dfrac{d_{31}^2}{s_{11}^E})\cdot w_p \cdot \dfrac{x_{Ri}-x_{Li}}{2h_p} \quad (18)$$

$\varepsilon_{33}^T$ is the permittivity of the piezoelectric material at constant stress. Using the assumed mode expansion method [32], the transverse displacement of the beam can be presented as

$$w = \sum_{j=1}^{Z}\phi_j(x)\eta_j(t) \quad (19)$$

where $\phi_j(x)$ is the $j$th normal mode function of a uniform clamped-clamped beam and $\eta_j(t)$ is the $j$th general coordinate. $Z$ is the number of modal functions we employ. Substituting Eq. (19) into Eqs.(13) and (17) and the lumped parameter model of the total system can be derived as

$$\begin{cases} \ddot{\boldsymbol{\eta}}+\mathbf{C}_d\cdot\dot{\boldsymbol{\eta}}+\boldsymbol{\Omega}\cdot\boldsymbol{\eta}-2\lambda\cdot\mathbf{G}\cdot\mathbf{U}=\mathbf{Q} \\ \mathbf{C}_p\cdot\dot{\mathbf{U}}-\mathbf{I}+\lambda\cdot\mathbf{G}^T\cdot\dot{\boldsymbol{\eta}}=0 \\ L\cdot\dot{\mathbf{I}}+R\cdot\mathbf{I}+P(\mathbf{U})=0 \end{cases} \quad (20)$$

where

$$\mathbf{C}_d = 2\xi\cdot diag(\omega_{n1},\omega_{n2}...\omega_{nZ}) \quad (21)$$

$$\boldsymbol{\Omega}=diag(\omega_{n1}^2,\omega_{n2}^2...\omega_{nZ}^2) \quad (22)$$

$$\mathbf{G}=\begin{bmatrix} \phi_1'(x_{R1})-\phi_1'(x_{L1}) & \cdots & \phi_1'(x_{RN})-\phi_1'(x_{LN}) \\ \vdots & \ddots & \vdots \\ \phi_Z'(x_{R1})-\phi_Z'(x_{L1}) & \cdots & \phi_Z'(x_{RN})-\phi_Z'(x_{LN}) \end{bmatrix} \quad (23)$$

$$\mathbf{Q}=F\cdot\begin{Bmatrix} \phi_1(x_F) \\ \vdots \\ \phi_Z(x_F) \end{Bmatrix} \quad (24)$$

$\mathbf{C}_d$ is the assumed damping matrix.

For the purpose of achieving nonreciprocal wave transmission, the asymmetry needs to be introduced and hence circuits in part I (Fig. 1) are not shorted. For this case, the modification is to add the dynamic equations of the linear *L-C-R* circuits to the electric part in Eq. (17).

## 3. Numerical simulations of wave transmission properties

With the derived dynamic equations of the proposed nonlinear piezo-metastructure in Eq.(20), we perform numerical investigations to study the wave transmission properties. We first investigate the local-resonance bandgap of the piezo-metastructure under small excitations, followed by analyzing the supratransmission phenomenon in the local-resonance bandgap. Then we explore the nonreciprocal wave transmission based on the supratransmission phenomenon.

In the numerical investigation, parameters of the bistable circuit are listed in Table 1 and parameters of the piezo-beam are shown in Table 2. The substrate-layer material is aluminium and the piezo-layer material is PZT-5H. We use the first 50 modal functions of a uniform clamped-clamped beam as trial functions to derive the approximate response of the piezo-beam. These trial functions are given in Appendix A. The harmonic force excitation $F = F_m \sin(2\pi f_e t)$ is applied to the system and the initial conditions for all dynamic variables are set to be the equilibrium values.

**Table 2** Parameters of the piezo-beam in simulation

| Parameters | Value |
|---|---|
| Mode function number $Z$ | 50 |
| Unit number $N$ | 16 |
| Beam total length $L_t$ | 909 [mm] |
| Actuating position $L_a$ | 10 [mm] |
| Sensing position $L_s$ | 40 [mm] |
| Unit length $p$ | 37 [mm] |
| Electrode length $l_e$ | 25 [mm] |
| Substrate-layer height $h_b$ | 6 [mm] |
| Piezo-layer height $h_p$ | 0.3 [mm] |
| Substrate-layer width $w_b$ | 5 [mm] |
| Piezo-layer width $w_p$ | 5 [mm] |
| Substrate-layer density $\rho_b$ | 2700 [Kg m$^{-3}$] |
| Piezo-layer density $\rho_p$ | 7800 [Kg m$^{-3}$] |
| Substrate-layer modulus $E_b$ | 69×10$^9$ [Pa] |
| Piezo-layer modulus $E_p$ | 72×10$^9$ [Pa] |
| Piezoelectric strain constant $d_{31}$ | -2.7×10$^{-10}$ [C m$^{-2}$] |
| Dielectric constant $e_{33}^T$ | 3.1×10$^{-8}$ [F m$^{-1}$] |
| Damping ratio $\xi$ | 0.02 |

### 3.1 Local-resonance bandgap

The bandgap is a frequency range within which the wave transmission is prohibited. In order to obtain the bandgap property of the proposed system, we excite the system on point $A_1$ and analyze the transmissibility $T$ from sensing point $S_1$ to $S_2$, which is defined as the ratio of root-mean-square value of velocity on point $S_2$ to that on point $S_1$, i.e. $T = rms(v_{S_2})/rms(v_{S_1})$.

For illustration purposes, a small excitation with forcing amplitude $F_m$=0.1 N is employed and the amplification factors of op-amps are set to be $g = 3, g_1 = 2$. Fig. 5 depicts the transmissibility versus frequency curves for cases with different $V_a$. It shows that, when the bistable circuits are shorted, there is no bandgap (solid black line); otherwise, the bandgap exists. This is because for the proposed nonlinear piezo-metastructure with small excitation, it can be approximated with a linear piezo-metastructure [19], which would exhibit the local-resonance bandgap. The observed local-resonance bandgap in Fig. 5 is around the calculated linearized resonance frequency $f_n$ of the bistable circuit using Eq. (12), which is 4600 Hz. This verifies the efficacy of the analytical model. Fig. 5 also shows that the bandgap shifts to a lower frequency range as $V_a$ decreases. This can be explained using Eq. (11) and (12), which indicates that when $V_a$ is decreased, the linearized resonance frequency $f_n$ will be decreased. These results indicate that we can achieve adaptive bandgap tuning by conveniently controlling the DC voltage source $V_a$.

Eq. (12) indicates that the linearized resonance frequency $f_n$ of the bistable circuit depends on $g$ and $g_1$. If the product of $g$ and $g_1$ is constant, $f_n$ is invariant and thus the local-resonance bandgap should be the same, which can be verified from the transmissibility curves shown in Fig. 6. In this case, the voltage $V_a$ is set to be 0 V, the force excitation amplitude is $F_m$=0.1 N and $gg_1$=6.

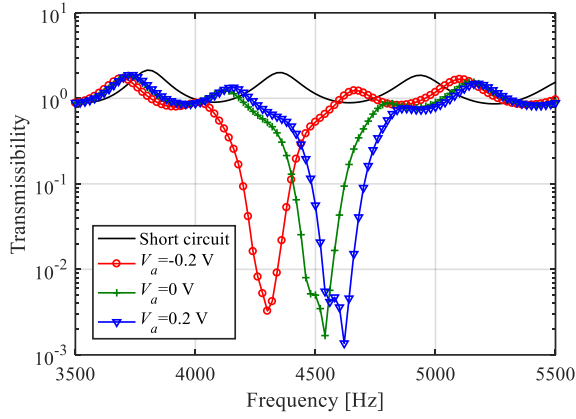

**Fig. 5** Effect of $V_a$ on the bandgap of the piezo-metastructure when $F_m$=0.1 N and $g=3, g_1=2$.

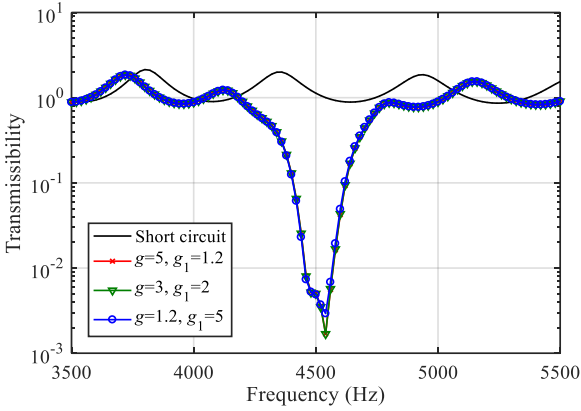

**Fig. 6** Effect of $g$ and $g_1$ on the bandgap of the piezo-metastructure when $F_m$=0.1 N, $V_a$=0 V and $gg_1=6$.

*3.2 Supratransmission*

Supratransmission [33] has been observed and analyzed in the mechanical bistable systems [14,34]. It refers to the phenomenon that when the excitation amplitude exceeds certain threshold value, large amplitude wave starts to transmit through the system even with input frequency inside the linearized bandgap. In this section, we will explore the supratransmission phenomenon in the local-resonance bandgap with the proposed bistable piezo-system.

From Section 3.1, we find that the proposed system exhibits local-resonance bandgap with small excitation force. In order to study the wave transmission properties with increasing forcing amplitudes, the relation between the output response $\bar{v}_o$ (sensing point $S_2$) and the excitation amplitude $F_m$ on $A_1$ is shown in Fig. 7, where three cases with different $V_a$ are simulated. $\bar{v}_o$ is defined as the ratio of root-mean-square of output velocity to that of the case when all circuits are shorted. Thus, when $\bar{v}_o$ is on the order of 1, it indicates that wave can propagate through the beam while $\bar{v}_o \ll 1$ means that wave transmission is prohibited. The excitation frequencies are fixed in the bandgap, which can be identified from Fig. 5 and are different for the three cases. Fig. 7 shows that the output response maintains at small values when the excitation is low, which is due to the effect of bandgap. However, supratransmission happens when the sufficiently large excitation is imposed, which triggers the snap-through of bistable circuits across the two stable equilibria. Fig. 5 demonstrates that varying $V_a$ could result in the shift of bandgap and Fig. 7 further reveals that for all three cases, supratransmission exists in the corresponding bandgap. Therefore, it can be concluded that adaptive frequency range for supratransmission is attainable in the proposed system by varying $V_a$.

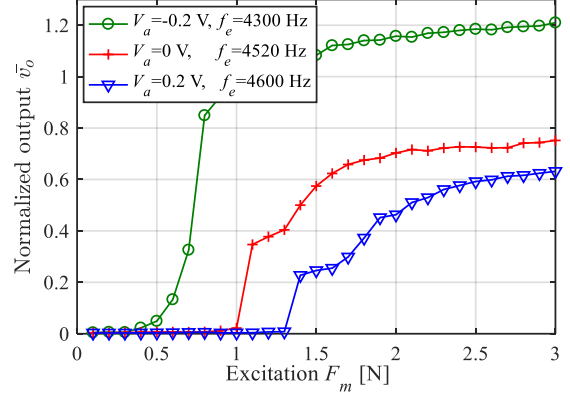

**Fig. 7** Supratransmission in three cases with different $V_a$ and $g=3, g_1=2$

Although adjusting $g$ and $g_1$ would not affect the local-resonance bandgap as along as $gg_1$ remains constant, as can be seen in Fig. 6, it may still influence the onset of supratransmission phenomenon since the stable equilibria of bistable circuits can be changed as shown in Fig. 4. Therefore, the effect of $g$ and $g_1$ on supratransmission is explored and illustrated in Fig. 8, where $g$ and $g_1$ are varied while $gg_1$ remains the same. It can be seen that the supratransmission threshold increases with increasing amplification factor $g$ (accordingly, decreasing $g_1$). This is because that when $g$ increases, the distance between the two stable equilibria of bistable circuits is increased, and thus a larger excitation is required to trigger snap-through of bistable circuits. These results indicate that for a prescribed local-resonance bandgap of the system with constant $gg_1$, we can still achieve adaptive supratransmission threshold by adjusting the ratio between $g$ and $g_1$.

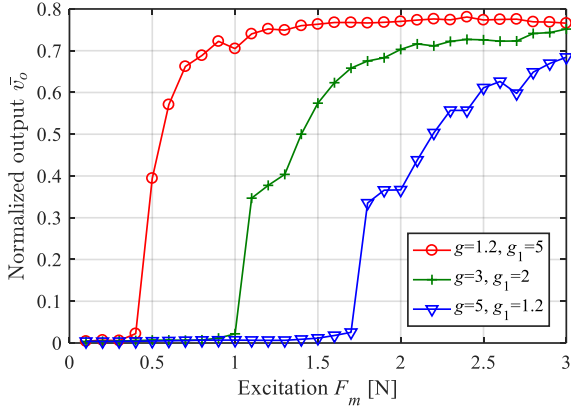

**Fig. 8** Effect of $g$ and $g_1$ on supratransmission when $gg_1$ is constant and $V_a$=0 V, $f_e$=4520 Hz

*3.3 Nonreciprocal wave transmission*

For a symmetric system, the supratransmission thresholds for wave propagating along the two opposite directions are the same. When asymmetry is introduced to intentionally engineer the supratransmission thresholds at different levels in the two transmission directions, one-way wave transmission would exist in the excitation range bounded by the two thresholds. To verify the hypothesis and illustrate the phenomenon, numerical simulations are performed with the proposed nonlinear piezo-metastructure.

For demonstration purposes, $V_a$ is set to be 0 V and the amplification factors are selected to be $g = 3, g_1 = 2$. The excitation frequency is 4520 Hz, which is in the bandgap of the linearized system. In contrast to short circuits on both ends of the system used in previous analysis on bandgap and supratransmission, linear *L-R* circuits are added on part I to intentionally induce asymmetry, as shown in Fig. 1(a). Since the linearized resonance frequency of the bistable circuits operating at the stable equilibria is 4600 Hz, to maintain the same resonance frequency of linear circuits in part I, the inductance $L_a$ is selected to be 0.112 H. The resistor $R_a$ in series with $L_a$ is chosen to be $R_a = 300\Omega, 150\Omega, 100\Omega$. When $R_a$ is decreased, the resonance effect of the linear *L-C-R* circuits would be magnified and thus the wave attenuation effect in part I is increased. Therefore, resistance $R_a$ in the *L-C-R* circuits are selected as a measure of asymmetry factor to evaluate the effect of asymmetry on the nonreciprocity of the proposed system.

The results shown in Fig. 9 illustrate that the supratransmission thresholds of waves in the forward direction are the same for all three cases and the thresholds of the backward wave direction are all greater than those of forward direction. In the excitation amplitude ranges identified by A, B and C, which are bounded by forward and backward supratransmission thresholds, we can clearly see that the forward transmission is large while the backward transmission is quite small. Hence, these are the nonreciprocity excitation amplitude ranges as the resistor $R_a$ varies. The phenomenon can be explained as follows. Since the resonance frequencies of linear circuits in part I have been adjusted to be the same with the linearized resonance frequency of the bistable circuit, wave propagating through part I would be attenuated if the excitation frequency is in the local-resonance bandgap. Therefore, when the beam is excited at point $A_1$ and wave propagates in the backward direction, wave amplitude would be attenuated in part I before entering into the part II where supratransmission happens, which means that the critical excitation amplitude at $A_1$ to trigger supratransmission would be larger compared to the case without part I. This point can be clearly seen by comparing supratransmission thresholds shown by lines with star markers of Fig. 9 with that shown by the line with cross markers of Fig. 7. However, when the beam is excited at point $A_2$ and wave propagates in the forward direction, the supratransmission threshold is not affected by part I. So the supratransmission thresholds of forward directions shown in Fig. 9 (lines with circle markers) remain the same with that shown by the line with cross markers of Fig. 7. Overall, this will result in that the supratransmission threshold of the backward direction is larger than that of the forward direction and thus creating an excitation range within which nonreciprocity appears.

Since decreasing $R_a$ could increase the wave attenuation effect in part I, it will increase the supratransmission thresholds in the backward direction and hence increase the nonreciprocity excitation amplitude range, but it will also reduce the forward transmission amplitude in this range, as can be seen in Fig. 9. Similar compromise between the nonreciprocal range of the excitation amplitude and the amplitude of the forward transmission was also observed in the nonlinear nonreciprocal optical system [35]. Hence, trade-off between them has to be considered when designing the system. Additionally, since the proposed nonreciprocity mechanism is based on supratransmission, which can be conveniently tuned by varying $V_a$ and/or the amplification factors of op-amps ($g$ and $g_1$) as explained in Sections 3.2, both the frequency range and excitation amplitude range to enable nonreciprocal wave transmission could be adaptively adjusted using the proposed system.

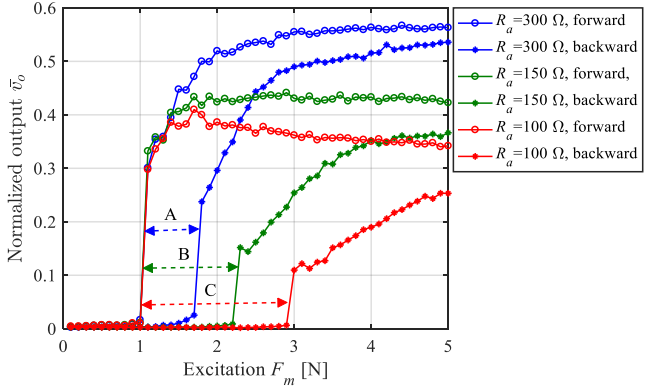

**Fig. 9** Nonreciprocal wave transmission in the nonlinear piezo-metastructure with different asymmetry factors and $V_a$=0 V, $g=3, g_1=2$. A, B and C are the excitation amplitude ranges with nonreciprocity when $R_a$ equals to 300 Ω, 150 Ω and 100 Ω, respectively.

## 4. Experimental investigation

Experiments are conducted to verify intriguing properties observed in numerical investigation and performance of the proposed nonlinear piezo-metastructure. The schematic and photograph of the experimental set-up are shown in Fig. 10 (a) and (b), respectively. The two piezo-patches $A_1$ and $A_2$ are used as actuators. The two piezo-patches $S_1$ and $S_2$ are employed as sensors, which are shunted with resistors $R_s$=15 KΩ. For backward direction, the excitation voltage is applied on $A_1$ and the voltage on sensing piezo-patch $S_2$ is measured as the output response; for forward direction, the excitation voltage is applied on $A_2$ and the voltage on $S_1$ is measured as the output response. Parameters of the piezo-beam are the same with that given in Table 2. The difference between the experimental piezo-beam and the one in the analytical model is that instead of building a bimorph piezo-beam with uniform piezo-layers bonded on both sides of the beam and periodic electrodes on the piezo-layers, which is difficult to realize practically, we attach piezo-patches periodically on both sides of the beam. This difference will not influence the qualitative behaviours of the system although the quantitative response would be different. Parameters of the experimental bistable circuits are given in Table 3. The diode type is 1N4148 and the op-amp type is TLE2141CP. The amplification factors of the op-amps with feedback can be calculated using parameters in Table 3 to be $g$=3, $g_1=2$. For the linear circuit in part I, we used synthetic inductors [36] as shown in Fig. 10(c) since the inductance $L_a$ of which can be easily adjusted by tuning the variable resistor $R_v$, as given by

$$L_a = \frac{R_5 R_7 R_v}{R_6} C \quad (25)$$

In the experiment, we tune $R_v$ to ensure that the resonance frequency of the linear circuit is the same with the linearized resonance frequency of the bistable circuit operating at its stable equilibria.

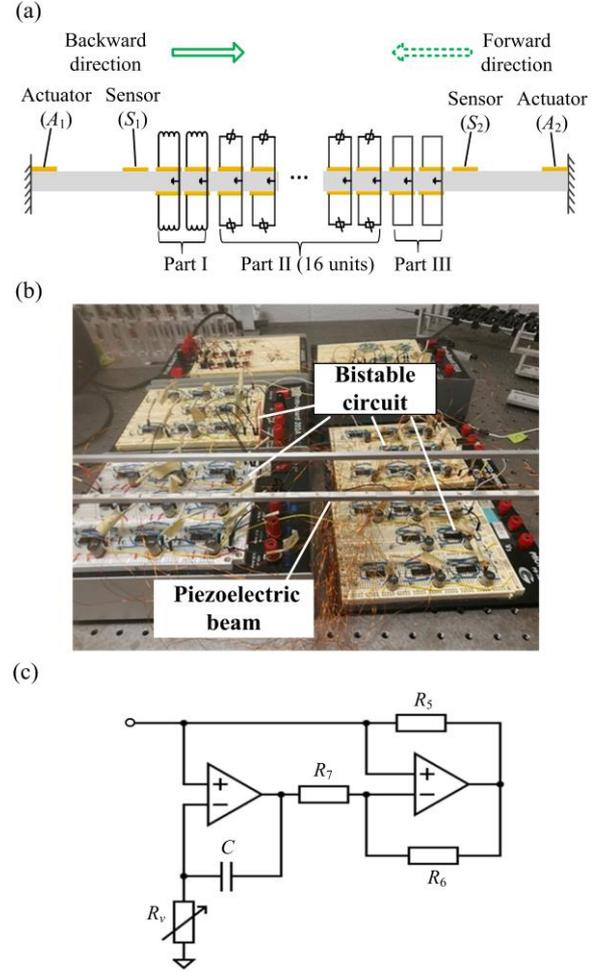

**Fig. 10** (a) Schematic of the experimental set-up; (b) photograph of the experimental set-up; (c) the circuit diagram of synthetic inductors used in the asymmetry part (part I).

**Table 3** Parameters of the bistable circuit in experiment

| Parameters | Value |
|---|---|
| $L$ | 0.1 [H] |
| $R$ | 87 [Ω] |
| $R_0$ | 15 [KΩ] |
| $R_1$ | 1.5 [KΩ] |
| $R_2$ | 3 [KΩ] |
| $R_3$ | 1.5 [KΩ] |

| $R_4$ | 1.5 [KΩ] |
|---|---|

The stable equilibria $U_e$ of the bistable circuit are measured with varying voltages $V_a$ applied by the DC voltage source. The results are shown in Fig. 11, which corroborate with previous finding that stable equilibria of the bistable circuit can be adjusted by changing $V_a$. Similar with the simulation results in Fig. 2, the experimental relation between $U_e$ and $V_a$ is quasi-linear.

To discover the bandgap property of the system, the linear circuits in part I are shorted. The excitation voltage is applied on piezo-patch $A_1$ and a frequency-sweeping voltage signal with amplitude $U_m$=2.2 V is used. The transmissibility from the voltage on sensing piezo-patch $S_1$ to the voltage on sensing piezo-patch $S_2$ is measured, as shown in Fig. 12. The results show that when the bistable circuits are shorted, there is no bandgap, otherwise, the bandgap exists. By adjusting the DC voltage source $V_a$, the bandgap shifts; a smaller $V_a$ will result in a lower frequency bandgap. Overall, the qualitative behaviours are the same with that shown in the simulation results, as can be seen in Fig. 5.

To verify the supratransmission phenomenon in the proposed system, we applied harmonic excitation voltage signal on $A_2$ with different amplitude $U_m$. $V_a$ is set as 0 V and the excitation frequency is chosen to be $f_e$=5150 Hz, which is in the bandgap, as shown in Fig. 12. The observed supratransmission phenomenon is shown in Fig. 13. $\bar{U}_o$ is defined as the ratio of root-mean-square of voltage $U_o$ on output sensing piezo-patch ($S_1$) to that obtained when all circuits are shorted. Hence, it indicates that wave can propagate when $\bar{U}_o$ is on the same order with 1 and wave cannot propagate when $\bar{U}_o$ is much smaller than 1. Since the attenuation in the bandgap is not quite large, as can be seen in Fig. 12, the output response under small excitation level in Fig. 13 is not as small as simulation results in Fig. 7. But we can still see clearly the sharp increase of output response after a certain excitation threshold.

Lastly, we seek to explore the nonreciprocal wave transmission experimentally. In order to study the effect of asymmetry factor on the nonreciprocity properties, we connected a resistor $R_b$ in series with the synthetic inductor used in part I, which has an inherent resistance [36] due to the non-ideality of practical op-amps, to change the total resistance $R_a$ of the L-C-R circuits used in the simulation and hence change the asymmetry factor. The experimental results are shown in Fig. 14, with two scenarios, $R_b$= 0 Ω and $R_b$=102 Ω. For both scenarios, the supratransmission thresholds of the backward direction are greater than those of the forward direction. Hence there exists the nonreciprocity excitation amplitude range. Comparing with the case $R_b$= 0 Ω, for the case $R_b$=102 Ω, the excitation amplitude range for nonreciprocity is reduced while the transmission amplitude in this range is increased, which is in accordance with the simulation results shown in Fig. 9.

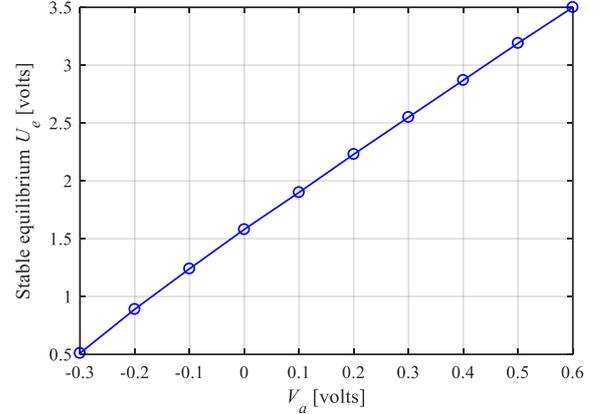

**Fig. 11** Experimental stable equilibria of the bistable circuit

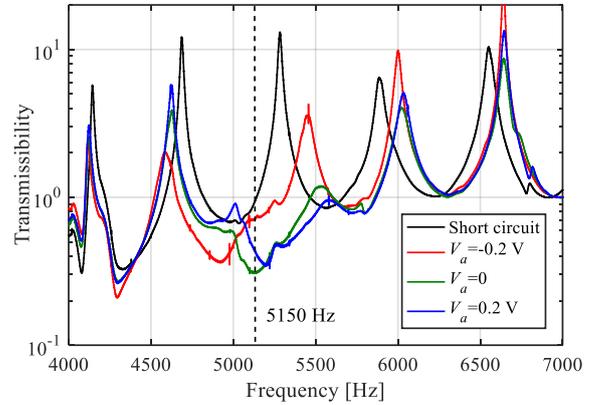

**Fig. 12** Experimental transmissibility under a small excitation voltage $U_m$=2.2 V.

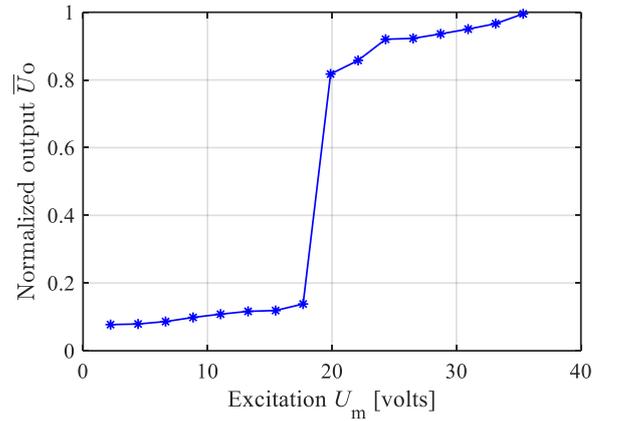

**Fig.13** Experimental supratransmission phenomenon when $V_a$=0 V, $f_e$=5150 Hz.

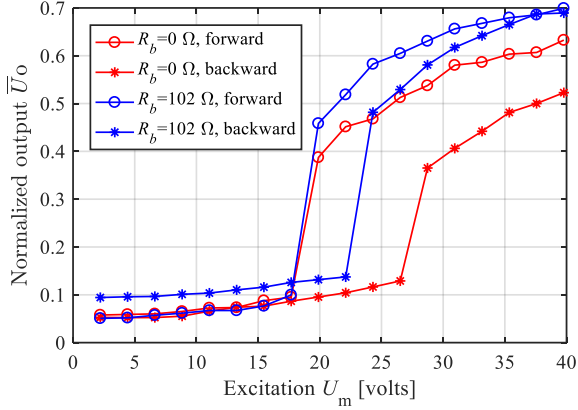

**Fig.14** Experimental nonreciprocal wave transmission when $V_a$=0 V, $f_e$=5150 Hz.

## 5. Conclusion

In this research, we propose a piezo-metastructure shunted with bistable circuits, and investigate numerically and experimentally the adaptive bandgap, supratransmission and nonreciprocal wave propagation phenomena in this system. Within the linearized local-resonance bandgap of the piezo-metastructure, we uncover the supratransmission phenomenon, a bandgap transmission property of nonlinear systems. It is then verified that by introducing circuit asymmetry, we are able to enable nonreciprocal wave energy transmission by triggering supratransmission at different input amplitudes in the two opposite directions. Additionally, we find that increasing the asymmetry factor could result in a larger excitation amplitude range to facilitate nonreciprocity while lowering the forward transmission amplitude in this range. Hence a trade-off between the breadth and the strength of the nonreciprocal response of the system has to be considered when selecting the asymmetry factor. Different from constituents of nonlinear mechanical metamaterials, the nonlinear properties of bistable circuits in the proposed piezo-metastructure could be easily adjusted by tuning the DC voltage sources or changing the amplification factors of op-amps with negative feedback. This feature can facilitate adaptive nonreciprocal wave transmission. Overall, the proposed system creates a paradigm shift for manipulating elastic wave transmission with unprecedented programmability, which would inspire more research activities in the field of nonlinear piezo-metastructures.

## ACKNOWLEDGEMENTS

This work is partially supported by the U.S. Army Research Office under grant number W911NF-15-1-0114, and by the National Science Foundation under Award No. 1661568. Yisheng Zheng also acknowledges the partial financial support from China Scholarship Council. The authors would like to thank Dr. Jinki Kim and Dr. Narayanan Kidambi for their useful discussions on the bistable circuit.

## Appendix A

The mode shape function of the uniform clamped-clamped beam is

$$\phi_j(x) = \cos(\beta_j x) - \mathrm{ch}(\beta_j x) + r \cdot [\sin(\beta_j x) - \mathrm{sh}(\beta x)] \quad (A.1)$$

where

$$r = \frac{\sin(\beta_j L_t) + \mathrm{sh}(\beta_j L_t)}{\cos(\beta_j L_t) - \mathrm{ch}(\beta_j L_t)} \quad (A.2)$$

and

$$\begin{cases} [\beta_1, \beta_2, \beta_3, \beta_4]^T = [4.730, 7.853, 10.996, 14.137]^T / L_t \\ \beta_j \approx (j+1/2)\pi / L_t, \text{if } j \geq 5 \end{cases} \quad (A.3)$$

If $j$>10, in order to avoid numerical issues [37], the shape function can be simplified as

$$\phi_j(x) \approx \cos(\beta_j x) - \sin(\beta_j x) - e^{(-\beta_j x)} + e^{(\beta_j x - \beta_j L_t)} \sin(\beta_j L_t) \quad (A.4)$$